\documentclass[12pt]{article}

\usepackage{graphicx}
\usepackage{amsmath}
\usepackage{amssymb}
\allowdisplaybreaks

\begin{document}

\baselineskip=17.5pt plus 0.2pt minus 0.1pt

\renewcommand{\theequation}{\arabic{equation}}
\renewcommand{\thefootnote}{\fnsymbol{footnote}}
\makeatletter
\def\CR{\nonumber \\}
\def\pt{\partial}
\def\be{\begin{equation}}
\def\ee{\end{equation}}
\def\bea{\begin{eqnarray}}
\def\eea{\end{eqnarray}}
\def\eq#1{(\ref{#1})}
\def\la{\langle}
\def\ra{\rangle}
\def\hyp{\hbox{-}}
\def\sixj#1#2#3#4#5#6{\left\{ \begin{matrix} #1&#2&#3 \\ #4&#5&#6 \end{matrix} \right\}}
\def\threej#1#2#3#4#5#6{\left( \begin{matrix} #1&#2&#3 \\ #4&#5&#6 \end{matrix} \right)}

\begin{titlepage}
\title{\hfill\parbox{4cm}{ \normalsize YITP-05-70 \\{\tt hep-th/0511154}}\\
\vspace{1cm} An invariant approach to dynamical fuzzy spaces \\ with a three-index variable
-- Euclidean models\footnote{
Submitted to the Proceedings of the 4th International Symposium
"Quantum Theory and Symmetries" (QTS-4), 15-21 August 2005, Varna, Bulgaria.
}}
\author{
Naoki {\sc Sasakura}\thanks{\tt sasakura@yukawa.kyoto-u.ac.jp}
\\[15pt]
{\it Yukawa Institute for Theoretical Physics, Kyoto University,}\\
{\it Kyoto 606-8502, Japan}}
\date{\normalsize November, 2005}
\maketitle
\thispagestyle{empty}

\begin{abstract}
\normalsize
A dynamical fuzzy space might be described in terms of a dynamical three-index variable ${C_{ab}}^c$, 
which determines
the algebraic relations $f_a f_b ={C_{ab}}^c f_c$ of the functions $f_a$ on a fuzzy space. 
A fuzzy analogue of the general coordinate transformation would be given by the general linear transformation 
on $f_a$. The solutions to the invariant equations of motion of ${C_{ab}}^c$ can be generally 
constructed from the invariant tensors of Lie groups.
Euclidean models the actions of which are bounded from below are introduced.
Lie group symmetric solutions to a class of Euclidean model are obtained.  
The analysis of the fluctuations around the SO(3) symmetric solution
shows that the solution can be regarded as a fuzzy $S^2/Z_2$.
\end{abstract}
\end{titlepage}

\section{Introduction}
\label{introduction}
Several thought experiments in semi-classical quantum gravity and string theory 
\cite{Garay}-\cite{salecker} imply uncertainties in 
space-time observables such as distances. This suggests that the smooth continuum space-time, 
which is the classical notion of space-time in general relativity, 
should be replaced with a quantum one in yet unknown theory of quantum gravity.
An interesting candidate is given by non-commutative geometry 
or fuzzy space \cite{Connes}. The main general idea of fuzzy space is that 
a space-time is not apriori given by coordinates, but is defined by the algebraic relations of the fields 
on it. Since matter and space-time
cannot be separately considered in physics, the main idea itself would be acceptable for many of 
the physicists, although it is harder to apply the idea to physics 
than the standard methods.

The dynamics of space-time can be classically described by the general relativity. 
On the other hand, it is not yet clear what can describe the dynamics of a fuzzy space-time
\footnote{The present author considered 
evolving fuzzy spaces in \cite{Sasakura:2003ke}. However, the formulation gives 
certain time-dependent fuzzy spaces, but does not give any dynamical fuzzy spaces influenced by
the back reactions from matter fields.}.
The description should approach the general relativity in classical limit.
The central principle of general relativity is the invariance under the general coordinate transformation.
It would be obviously interesting if a dynamical fuzzy space can be obtained from formulation invariant 
under a fuzzy analog of the general coordinate transformation.
This kind of formulation would lead to fuzzy general relativity, 
which may describe dynamical fuzzy spaces.

A fuzzy space is characterized by the algebraic relations $f_a f_b={C_{ab}}^c f_c$ of
the functions $f_a$ on it.
Therefore it would be a natural assumption that a dynamical fuzzy space can be obtained by
making the three-index variable ${C_{ab}}^c$ a dynamical variable. 
It will be argued in the following section that
a fuzzy analog of the general coordinate transformation is given by  
the general linear transformation on $f_a$. Thus the central proposal 
of this talk, which is mainly based on the paper \cite{Sasakura:tensor},
is that fuzzy space dynamics should be described by a theory of dynamical ${C_{ab}}^c$ 
with the invariance under the general linear transformation.

The above proposal is that dynamical fuzzy spaces are described by a kind of tensor models. 
Actually tensor models appeared previously in the study of quantum gravity as
a description of the simplicial quantum gravity in more than two dimensions \cite{tensor}.
However there are differences in the treatment of the tensor models between the simplicial quantum gravity 
and the dynamical fuzzy spaces. 
In the context of the simplicial quantum gravity, 
the integration with the tensor variables generates the partition function,
and the classical solutions are not important. 
On the contrary, semi-classical treatment plays central roles in this talk.
In other words,
it is assumed that the tensor models can be treated semi-classically in certain limits.
Another difference from the simplicial gravity is that tensors of rank higher than three
are not considered irrespective of the dimensions of space. 
I expect that fuzzy spaces of various dimensions can be generated from three-tensor models. 
This will be an appropriate feature of the models,
because the dimensions should be a dynamical effective quantity of a fuzzy space. 

In the following section, a fuzzy analog of the general coordinate 
transformation is discussed and the general form of the models is proposed.  
In Section \ref{thesolutions}, it is shown that the solutions to the equations of motion 
can be generally constructed from the invariant tensors of Lie groups. 
In Section \ref{euclideanmodel}, Euclidean models are introduced, the actions of which are bounded from below.
In Section \ref{solution}, the Lie group symmetric classical solutions to a class of Euclidean models
are obtained.
In Section \ref{fluctuation}, the fluctuations around the SO(3) symmetric solution are analyzed
in the quadratic order,
and the space associated with the solution is identified with fuzzy $S^2/Z_2$. 
The final section is devoted to discussions and comments.

\section{A fuzzy general coordinate transformation and the general form of the models}
\label{themodel}
Let me first review the general coordinate transformation in the usual continuous space $R^d$. 
A basis of the continuous functions on $R^d$ can be given by 
\begin{equation}
\{ 1, x^i, x^ix^j, x^ix^jx^k,\cdots \},
\end{equation}
where $x^i\ (i=1,\cdots,d)$ are the coordinates of $R^d$. Let $f_a$ with an index $a$ denote these independent 
functions in the set.
A continuous function on $R^d$ is given by a linear combination of $f_a$.

Let me consider a general coordinate transformation,
\begin{equation}
\label{coordtrans}
{x'}^i=t^i(x^1,\cdots,x^d).
\end{equation}
It is a natural restriction that the coordinate transformation is not singular, and is invertible.
Since the right-hand side is a continuous function, the general coordinate transformation can be
rewritten as a linear transformation,
\begin{equation}
\label{transx}
{x'}^i={M_i}^a f_a,
\end{equation}
where ${M_i}^a$ are real.

Now let me consider a fuzzy space with a finite number of independent functions $f_a\  (a=1,\cdots,n)$
on it. This fuzzy space is a fuzzy analogue of a compact space in usual continuous description.   
It is  assumed that all the $f_a$ are real functions, 
and therefore the variable ${C_{ab}}^c$, which determines the algebraic relations
$f_a f_b ={C_{ab}}^c f_c$, is assumed to be real. 
Neither the associativity nor the commutativity of the 
algebra is assumed, so that ${C_{ab}}^c$ has no other constraints.

An important assumption of this talk is to interpret the transformation rule \eq{transx} 
as a partial appearance of a more general linear transformation,
\begin{equation}
\label{transm}
f'_a={M_a}^b f_b,
\end{equation}
where ${M_a}^b$ can take any real values provided that the matrix ${M_a}^b$ is invertible, which comes 
from the assumed invertibility  of the coordinate transformation \eq{coordtrans}. 
Thus the fuzzy general coordinate transformation is the GL($n$,$R$) transformation on $f_a$.
Under the transformation \eq{transm}, the three-index variable ${C_{ab}}^c$ transforms in 
\begin{equation}
\label{transc}
{{C'_{ab}}^c}={M_a}^{a'}{M_b}^{b'} {C_{a'b'}}^{c'}{ (M^{-1})_{c'}}^c.
\end{equation}

In the next, let me discuss the construction of an action invariant under the GL($n$,$R$) transformation.
From the transformation property \eq{transc}, the 
lower and upper indices must be contracted to make an invariant.
Since ${C_{ab}}^c$ has more lower indices  
than the upper, it is necessary to introduce an additional variable which has more upper indices to 
construct an invariant action as a polynomial function of ${C_{ab}}^c$.    
Therefore I introduce a new variable with two upper indices, $g^{ab}$, which is real and symmetric in the 
indices, $g^{ab}=g^{ba}$.
Thus an invariant action is a function of the two dynamical real variables $g^{ab}$ and ${C_{ab}}^c$,
\begin{equation}
\label{generalaction}
S(g^{ab}, {C_{ab}}^c),
\end{equation}
where all the indices are contracted. 

\section{Classical solutions and Lie group symmetries}
\label{thesolutions}
In this section I discuss the general properties of the classical solution to the equation of motion
from the invariant action \eq{generalaction}. 
Let me suppose that the equation of motion of ${C_{ab}}^c$, 
\be
\label{eomgeneralofc}
\frac{\partial S}{\partial {C_{ab}}^c}= 0,
\ee
is satisfied for ${C_{ab}}^c={C^0_{ab}}^c,\ g^{ab}=g_0^{ab}$.  It can be shown \cite{Sasakura:tensor} 
that, using the invariance under the GL($n$,$R$) transformation,
the equation of motion of $g^{ab}$ is automatically satisfied
if $g_0^{ab}$ is not degenerate and \eq{eomgeneralofc} is satisfied.

Now let me consider a Lie group which has a representation of dimension $n$. The representation can be 
either reducible or irreducible, and can be either 
real or even complex, provided that the invariant tensors considered below are real.  
Let me embed the representation to a classical solution: 
The lower index of a classical solution is assumed to be transformed in 
the representation, while the upper one in the inverse representation.

Let $g_0^{ab}$ be a real symmetric rank-two invariant tensor under the inverse representation. 
I assume the tensor $g_0^{ab}$ is invertible as a matrix. 
Let me introduce ${I^\alpha_{ab}}^c\ (\alpha=1,2,\cdots,N)$, 
which span all the real tensors invariant under the same transformation property as ${C^0_{ab}}^c$.
Let me consider an ansatz that ${C^0_{ab}}^c$ is given by a linear combination of these invariant tensors,
\be
\label{caexpression}
{C^0_{ab}}^c= A_\alpha {I^\alpha_{ab}}^c,
\ee
where $A_\alpha\ (\alpha=1,2,\cdots,N)$ are real coefficients.
Since the action is obviously invariant under the transformation of the Lie group, 
the left-hand side of \eq{eomgeneralofc} becomes
a real invariant tensor when $g^{ab}$ and ${C_{ab}}^c$ are substituted with the invariant tensors above.
Therefore, after exchanging the upper and lower indices by using $g_0^{ab}$ and its inverse $g^0_{ab}$,
the left-hand side of \eq{eomgeneralofc} can be expanded with
${I^\alpha_{ab}}^c$,
\be
g^0_{ai} g^0_{bj} g_0^{ck} \left.\frac{\partial S}{\partial {C_{ij}}^k}
\right|_{\renewcommand\arraystretch{0.5} \begin{matrix} \scriptstyle g=g_0 \\ \scriptstyle C=C^0 \end{matrix}}=
 B_\alpha(A_1,A_2,\cdots,A_N){I^\alpha_{ab}}^c,
\ee
where $B_\alpha(A_1,A_2,\cdots,A_N)$ are functions of $A_1,A_2,\cdots,A_N$, which are
determined from the form of $S$.
Therefore the equation of motion \eq{eomgeneralofc} is reduced to the following
simultaneous equations for $A_1,A_2,\cdots,A_N$,
\be
\label{simbeq}
B_\alpha(A_1,A_2,\cdots,A_N)=0,\ \  (\alpha=1,\cdots,N).
\ee  
These equations are generally much easier to solve than \eq{eomgeneralofc}.
Since the numbers of the variables and the equations are the same, the simultaneous equations \eq{simbeq}
will generally have some number of solutions. The solutions are generally complex, 
but real solutions can be actually found in some interesting cases.  Some series of SO(3) invariant
solutions to some examples of equations of motion are given in \cite{Sasakura:tensor}.


\section{Euclidean models}
\label{euclideanmodel}
In the following I consider the constrained models which are expected to 
produce fuzzy spaces with Euclidean signatures. 
It is imposed that
\be
\label{gdelta}
g^{ab}=\delta_{ab},
\ee
and $g^{ab}$ is not dynamical. Even though the underlying symmetry has been changed
from GL($n,R$) to SO($n$), 
this Euclidean model seems to be interesting enough. 
One reason is that, as mentioned in the previous section, the detailed value of $g^{ab}$ is not important 
in constructing the classical solutions.
Another reason is that, while the number of the degrees of freedom of ${C_{ab}}^c$
is $O(n^3)$, that of $g^{ab}$ is only $O(n^2)$, and may be negligible
in the continuum limit $n\rightarrow\infty$, regardless of whether $g^{ab}$ is dynamical or not.
An advantage of the constraint \eq{gdelta} is that one can construct an action bounded from below.
The existence of a bound will make it easier to find a stable classical solution.
 
To construct a bounded action, let me start with an elementary fact that
a real matrix $M$ identically satisfies the following inequality,
\be
{\rm Tr}(M^TM) \geq 0,
\ee 
where Tr and $T$ denotes the trace and the transpose of matrix, respectively.
Consider real matrices $M_a \ (a=1,2,\cdots,n)$, and apply this inequality to $M= M_a +  M_a^T$.
Then one obtains
\be
\label{mtm}
{\rm Tr}(M_a M_a) + {\rm Tr}(M_a^T M_a) \geq 0, 
\ee
where the repeated index $a$ is summed over, and the same abbreviation will be implicitly assumed below.
In the same way,
another inequality can be obtained from $M= M_a^T M_b- M_b M_a^T$,
\be
\label{cross}
{\rm Tr}(M_a M_a^T M_b M_b^T + M_a^T M_a M_b^T M_b)-2 {\rm Tr}(M_aM_bM_a^TM_b^T)\geq 0.
\ee 

The following quantities are bounded from below,
\bea
\label{4to2}
&& c\, {\rm Tr}(M_a^TM_a)+ {\rm Tr}((M_a^T M_a)^2),  \cr
&& c\, {\rm Tr}(M_aM_a^T)+ {\rm Tr}((M_a M_a^T)^2),
\eea  
where $c$ is an arbitrary real number. This can be easily proven in the basis which 
diagonalizes $M_a^TM_a$ or $M_aM_a^T$, respectively.
From \eq{mtm}, \eq{cross} and \eq{4to2}, it can be easily shown that the following quantity is bounded 
from below,
\bea
\label{quantbound}
S_M&=&
{\rm Tr}(M_a M_a) +g_1 {\rm Tr}(M_a M_a^T M_b M_b^T +M_a^T M_aM_b^T M_b )\cr
&&\hspace{4cm} -2 g_2 {\rm Tr}(M_aM_bM_a^TM_b^T),
\eea
provided
\be
\label{g12}
g_1 > g_2.
\ee

\begin{figure}
\begin{center}
\includegraphics[width=130mm]{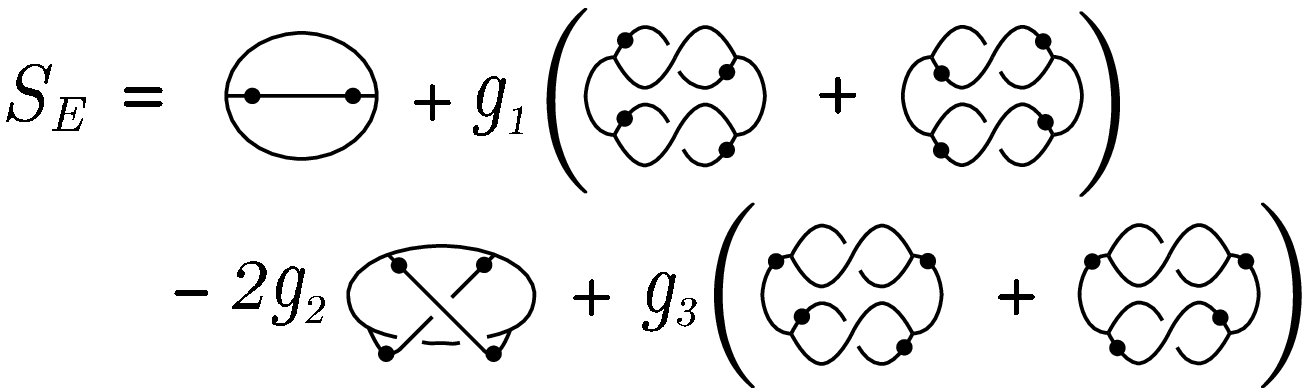}
\end{center}
\caption{The Euclidean action in the discussions. It is bounded from below when $g_1>g_2$ and $g_3\geq 0$.
The three vertices represent $C_{abc}$, the dots indicate the first index, and the lines 
represent the contractions of the indices.}
\label{fig1}
\end{figure}

Substituting $(M_a)_{bc}=C_{abc}$ into \eq{quantbound}, one obtains a function of $C_{abc}$, which
is bounded from below and can be graphically represented as the sum of the first and the $g_1$ and $g_2$ terms 
of $S_E$ in Fig.\ref{fig1}.
In the figure, the three vertices represent $C_{abc}$, the dots indicate the first index, and the lines 
represent the contractions of the indices. 
The $g_3$ term of $S_E$ in Fig.\ref{fig1} is also non-negative if
\be
\label{g3}
g_3\geq 0,
\ee
because 
\bea
&&C_{abc}C_{dbc}C_{eaf}C_{edf}+C_{abc}C_{dbc}C_{efa}C_{efd} \cr
&& \hspace{.5cm}=\sum_{bcef}\Big(\sum_a C_{abc}C_{eaf}\Big)^2+\Big(\sum_aC_{abc}C_{efa}\Big)^2\geq 0. 
\eea
Thus the action $S_E$ given graphically in Fig.\ref{fig1} is bounded from below, 
when \eq{g12} and \eq{g3} are satisfied. 

\section{The solutions to the Euclidean model}
\label{solution}
In this section, I will obtain the solutions to the equation of motion derived from the bounded action 
$S_E$ in Fig.\ref{fig1}. Let me first discuss the solutions to the equation of 
motion derived from $S_M$ in \eq{quantbound}, which is a part of the total action $S_E$. In due
course I will show that the solutions found below for $S_M$ actually satisfy the equation of motion from
the total action $S_E$. 

Let me start with a Lie group $G$ which has a finite dimensional real representation $R$. 
It is assumed also that the representation is irreducible and unitary. The real generators can be assumed to satisfy
\be
\label{prop1}
\begin{array}{rcl}
J_i^T & = & -J_i, \\
{[}J_i,J_j]&=&f_{ijk} J_k, \\
 J_i J_i &=& -c_1 I, 
\end{array}
\ee
where $I$ is an identity matrix, $c_1$ is a real number, and 
$f_{ijk}$ is the structure constant antisymmetric in the indices and satisfies
\be
\label{prop2}
f_{ijk}f_{i'jk}=c_2 \delta_{ii'},
\ee
where $c_2$ is a real number.

Now let me denote the matrix indices of the generators by the capital roman indices like $(J_i)_{AB}$.
The index of the matrices $A$ (and $B$) ranges from 1 to dim($R$), while the index of the generators $i$ 
ranges from 1 to dim($G$). Let me combine the two kinds of indices into one index, and identify it with 
the indices of the tensor $C_{abc}$. Namely,
\be
\begin{array}{ll}
A=a & {\rm for}\  a=1,\cdots, {\rm dim}(R), \\
\ i= a-{\rm dim}(R) & {\rm for}\  a={\rm dim}(R)+1,\cdots, {\rm dim}(R)+{\rm dim}(G).
\end{array}
\ee
Then the ansatz of the solution is given by  
\be
\label{ansatzofc}
C_{abc}=\left\{
\begin{array}{cl} k\,
(\, J_{a-{\rm dim}(R)})_{bc} & {\rm for}\  
\begin{array}{l}
{\rm dim}(R)<a\leq {\rm dim}(R)+{\rm dim}(G)\\
1\leq b,c \leq {\rm dim}(R) 
\end{array}, \\
0 & {\rm otherwise},
\end{array}
\right. 
\ee
where $k$ is a non-zero constant to be determined. 

The equation of motion from $S_M$ is given by
\be
\label{eqmforsm}
M_a^T+2 g_1 (M_a M_b^TM_b+M^bM_b^TM_a)-2g_2(M_b^TM_aM_b+M_bM_aM_b^T)=0.
\ee 
Substituting the ansatz \eq{ansatzofc} into the equation of motion through $(M_a)_{bc}=C_{abc}$ 
and using the properties 
\eq{prop1}, \eq{prop2}, one can show that \eq{ansatzofc} is a solution to \eq{eqmforsm} if
\be
\label{valuec}
k=\pm \frac1{\sqrt{4(g_1-g_2) c_1+2 g_2 c_2}}.
\ee

As for the $g_3$ term of $S_E$, each term contains two contractions between the first indices and the second or
the third ones. When the equation of motion of $C_{abc}$ from the total action $S_E$ 
is written down, the contributions from the $g_3$ term contains at least one contraction between 
the first index and the second or the third one. These contributions vanish,
because the tensor $C_{abc}$ of the ansatz \eq{ansatzofc} can take non-zero values only when 
the first index and the second or the third one are in the distinct ranges. 
 
\section{Analysis of the fluctuations around the SO(3) symmetric solution}
\label{fluctuation}
In this section, I study the fluctuations around the solution
\eq{ansatzofc}, \eq{valuec} with $G$=SO(3) in the quadratic order.
The real irreducible unitary representations of SO(3) can be classified by an integer spin $j$. 
The primary motivation for the analysis in this section is that the low frequency spectra of the fluctuations
will define the effective continuum space generated from the solution. It will be found that 
the solution with $G$=SO(3) can be identified as a fuzzy $S^2/Z_2$. Another motivation is that 
the signs of the eigenvalues of the quadratic fluctuations determine whether a solution is locally stable or not. 
It will be found that, in a broad range of the parameters $g_{1,2,3},j$, 
all the eigenvalues are positive except the Nambu-Goldstone-like zero
modes from the symmetry breaking of SO($n$) to SO(3).

Let me start with a coordinate change. In the analysis, I want to use the 
properties of the invariant tensors of SO(3) \cite{3jidentities}. The properties are usually expressed
in the basis of $J_3$ eigenvectors, but the invariant two-tensor in such a basis has the form 
$g^{m_1m_2} \propto \delta_{m_1,-m_2}$, which has a form different from \eq{gdelta} of the Euclidean model. 
A linear transformation from the $J_3$ eigen basis to the Euclidean basis can be given by
\be
\label{trans}
\begin{array}{rcl}
\left(
\begin{array}{c}
(j,m,+)_E \\
(j,m,-)_E
\end{array}
\right)
&=& \frac{i^{j+m}}{\sqrt{2}}  
\left( 
\begin{array}{cc}
1 &1 \\
i & -i 
\end{array}
\right)
\left(
\begin{array}{c}
(j,m) \\
(j,-m)
\end{array}
\right)
\ \ {\rm for}\ m>0,  \\
(j,0)_E&=& i^j (j,0)\ \ {\rm for}\ m=0. 
\end{array}
\ee
Here $(j,m,\pm)_E,(j,0)_E$  are the coordinates in the Euclidean model, while $(j,m)$ are the coordinates 
in the $J_3$ eigen basis. The two-tensor \eq{gdelta} of the Euclidean model is transformed by \eq{trans} to
\be
\label{twotensor}
g^{(j,m)\ (j',m')}=g_j^{mm'}\delta_{jj'}=(-1)^{j+m} \delta_{m,-m'}\delta_{jj'},
\ee
where I have used $(-1)^m=(-1)^{-m}$ for an integer $j$. In fact this is an invariant two-tensor in the 
$J_3$ eigen basis. 

In the next, I discuss the implications of the reality condition on the tensor $C_{abc}$ of the 
Euclidean model in the $J_3$ eigen basis. 
Let me first discuss the transformations of the invariant tensors. 
The invariant three-tensor of SO(3) is given by the $3j$-symbol, 
$\left(\begin{array}{c} j_1\ \ \ j_2\ \ \ j_3 \\ m_1\ m_2 \ m_3\end{array}\right)$ in the $J_3$ eigen basis. 
In the Euclidean basis, this becomes
\bea
\left(
\begin{array}{ccc}
j_1 & j_2 & j_3 \\
m_1 & m_2 & m_3 \\
+ & + & +
\end{array}
\right) \hspace{7cm} \cr
= \frac{i^{\sum_k j_k+m_k}}{2^{3/2}}\left\{ 
\left(\begin{array}{ccc}j_1& j_2 &j_3 \\ m_1& m_2 &-m_3\end{array}\right)
+
\left(\begin{array}{ccc}j_1& j_2 &j_3 \\ -m_1& -m_2 & m_3\end{array}\right)
\right\},
\eea
where the case $m_1+m_2-m_3=0\ (m_{1,2,3}> 0)$ is considered as an example.
The right-hand side is real, because of the conservation of the angular momentum $m_1+m_2-m_3=0$, 
the property $\left(\begin{array}{ccc}j_1 & j_2 & j_3 \\ -m_1&  -m_2 & m_3\end{array}\right)=
(-1)^{j_1+j_2+j_3}
\left(\begin{array}{ccc}j_1& j_2 &j_3 \\ m_1& m_2 & -m_3\end{array}\right)$, and the reality
of the $3j$-symbol. The same reality in the Euclidean basis can be shown for all the other components.
This implies that the ansatz for the classical solution can be assumed to be in the form 
(real number) $\times$ ($3j$-symbol) in the $J_3$ eigen basis. 
As for the solution given in Section \ref{solution} for $G$=SO(3), the index space of the three-tensor
is composed of the spin 1 and the spin $j$ representation spaces. The classical solution $C^0_{abc}$ 
can take non-zero values only when the first and the other indices are in the spin 1 and spin $j$ representation 
spaces, respectively. It is easy to determine the numerical factor, and one obtains the non-trivial 
components of the classical solution as
\be
\label{threetensor}
C^0_{(1,m_1)\ (j,m_2)\ (j,m_3)}=\pm \sqrt{\frac{j(j+1)(2j+1)}{4(g_1-g_2)j(j+1)+4g_2}}  
\left(\begin{array}{ccc}1& j &j  \\ m_1& m_2 & m_3\end{array}\right)
\ee
in the $J_3$ eigen basis, while the other components vanish.

\def\threej#1#2#3#4#5#6{\left(\begin{array}{ccc} #1 & #2 & #3 \\ #4 & #5 & #6 \end{array}\right)}
Since the classical solution is symmetric under SO(3), the fluctuations from the classical
solution can be classified by the SO(3) transformation properties.  
Namely, the classification can be done by the spins of the 
three indices $a,b,c,$ as well as the total spin $f$ of the fluctuation and the composite spin $h$  
of $b$ and $c$. More explicitly, the general form of the fluctuations can be 
written as 
\be
\label{flc}
\begin{array}{l}
\delta C_{(j_1,m_1)\ (j_2,m_2)\ (j_2,m_3)}=\sqrt{(2h+1)(2f+1)} 
\sum_{h,f}  g_{h}^{-m_2-m_3\ m_2+m_3} g_f^{-m\ m} \\
\times \threej{j_1}{h}{f}{m_1}{m_2+m_3}{-m} 
\threej{h}{j_2}{j_3}{-m_2-m_3}{m_2}{m_3}\delta C_{\alpha,f,m},
\end{array}
\ee
where $\alpha=\{j_1,j_2,j_3,h\}$, $m=m_1+m_2+m_3$, and $j_{1,2,3}$ take $1$ or $j$. 
The factor in front is determined from 
the normalization condition $g^{aa'}g^{bb'}g^{cc'} \delta C_{abc} \delta C_{a'b'c'}=1$.
Note that, because of the triangle inequalities of the composition of spins, 
the summation over $h$ and $f$ are finite. 
Because of the SO(3) symmetry, the action at the quadratic order of the fluctuations should have the form,
\be
\label{quadraticpart}
S_E^{(2)}=\sum_{\alpha,\alpha',f,m} s_2(\alpha,\alpha',f)\ g_f^{m,-m} 
\delta C_{\alpha,f,m} \delta C_{\alpha',f,-m}.
\ee 
In the Euclidean basis, this implies that the eigenvalues of the quadratic fluctuations can be 
obtained by diagonalizing the matrix $s_2(\alpha,\alpha',f)$ with respect to the indices $\alpha,\alpha'$.
Each eigenvalue has degeneracy $2f+1$.

Obtaining the matrix element $s_2(\alpha,\alpha',f)$ is a straightforward but a rather tedious work.
This work can be a little bit simplified by considering the trace over the momentum index $m$.
Namely, $\delta C_{\alpha,f,m}\delta C_{\alpha',f,-m}$ in \eq{quadraticpart} 
is replaced with $\delta_{\alpha \alpha_0} \delta_{\alpha' \alpha'_0} \delta_{ff_0} (g_{f_0}^{-m\ m})^{-1}$.
Then $S_E^{(2)}$ in \eq{quadraticpart} 
becomes $(2f_0+1) s_2(\alpha_0,\alpha'_0,f_0)$, where the factor in front comes 
from the summation over the momentum $m$.  Therefore, one can obtain $s_2(\alpha_0,\alpha'_0,f_0)$
by evaluating the quadratic order of the fluctuations of $\frac{1}{2f_0+1} S_E$ with the substitution
of \eq{flc} and the replacement of $\delta C_{\alpha,f,m}\delta C_{\alpha',f,-m}$ in the way above. 
The factor $\frac{1}{2f_0+1}$ cancels the
part $\sqrt{2f+1}$ of the normalization in \eq{flc} in the quadratic order.

\begin{figure}
\begin{center}
\includegraphics[width=30mm]{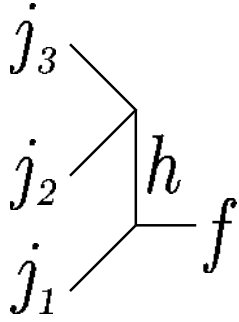}
\end{center}
\caption{The graphical representation of the fluctuation \eq{flc}. The normalization factor $\sqrt{2h+1}$
is implicit.
Each vertex represents the $3j$-symbol. Each line between vertices represents the contraction of the
indices with the two-tensor \eq{twotensor}. }
\label{fig2}
\end{figure}

\begin{figure}
\begin{center}
\includegraphics[width=50mm]{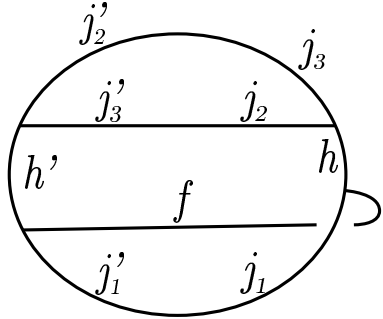}
\end{center}
\caption{The graphical representation of the contribution to $s_2(\alpha,\alpha',f)$
from the first term of $S_E$ in Fig.\ref{fig1}.
The implicit normalization factor is $\sqrt{(2h+1)(2h'+1)}$.}
\label{fig3}
\end{figure}

\def\sixj#1#2#3#4#5#6{\left\{\begin{array}{ccc} #1 & #2 & #3 \\ #4 & #5 & #6 \end{array}\right\}}
The graphical representation is convenient for the explicit computation. 
The fluctuation \eq{flc} can be represented as in Fig.\ref{fig2}.
Then the procedure of evaluating the quadratic order explained in the previous paragraph
can be graphically rephrased as that every choice of two vertices in the action of Fig.\ref{fig1} are 
replaced with the fluctuation Fig.\ref{fig2}, and the lines with $f$ are connected. 
For example, the contribution 
from the first term of Fig.\ref{fig1} becomes Fig.\ref{fig3} after the replacement. 
This contribution can be computed explicitly by using the identities in \cite{3jidentities}. 
Including also the implicit normalization factor $\sqrt{(2h+1)(2h'+1)}$, 
the contribution from the first term to $s_2(\alpha,\alpha',f)$ is evaluated as 
\be
(-1)^{h+j_2+j_3} \delta_{j_1j_1'}\delta_{j_2j_3'}\delta_{j_3j_2'}\delta_{hh'}.
\ee 
The same procedure can be done for all the contributions, and the final expression is given by
\be
\begin{array}{l}
s_2(\alpha,\alpha',f)= 
(-1)^{h+j_2+j_3} \delta_{j_1j_1'}\delta_{j_2j_3'}\delta_{j_3j_2'}\delta_{hh'} \\
+g_1 B \delta_{j_1j_1'}\delta_{j_2j_2'}\delta_{j_3j_3'} \Bigg[ \frac2{2j+1} \delta_{hh'}(\delta_{j_2 j}
+\delta_{j_3j})\\ 
+2 D (-1)^f(1+(-1)^{h+h'})\sixj{j}{j}{f}{1}{h}{j} \sixj{j}{j}{f}{1}{h'}{j} \delta_{j_11}\delta_{j_2j}\delta_{j_3j} \\ 
+2D (-1)^{h+h'} \sixj{j_2}{j}{f}{1}{h}{j} \sixj{j_2}{j}{f}{1}{h'}{j} \delta_{j_11}\delta_{j_3j} \\ 
+2D \sixj{j_3}{j}{f}{1}{h}{j} \sixj{j_3}{j}{f}{1}{h'}{j} \delta_{j_11}\delta_{j_2j} \Bigg] \\
-2 g_2 B D(-1)^{j+1} \delta_{j_11} \delta_{j_1'1} \delta_{j_2j_3'}\delta_{j_3j_2'}
[(-1)^{j_3+h'}E(j_3) \delta_{j_2 j}+(-1)^{j_2+h}E(j_2)\delta_{j_3j}] \\
-2g_2 B \delta_{j_1j_1'}\delta_{j_2j_2'}\delta_{j_3j_3'} \Bigg[ -2(-1)^h
 \sixj{j}{j}{h}{j}{j}{1} \delta_{hh'}\delta_{j_2j}\delta_{j_3j} \\
+ D E(j_3) \delta_{j_1 1}\delta_{j_2 j}+(-1)^{h+h'} D E(j_2) \delta_{j_11}\delta_{j_3j} \Bigg] \\
+g_3 B \delta_{hh'}\delta_{j_1j_1'}\delta_{j_2j_2'}\delta_{j_3j_3'} \Big[ \frac{2}{2j+1} \delta_{j_1j}
+\frac13 \delta_{j_21}+\frac13 \delta_{j_31}\Big]\\
+g_3 B (-1)^f \\
\times \Bigg[ \sqrt{\frac{2h'+1}{3}} \sixj{j}{f}{1}{h'}{j}{1}\big( (-1)^j \delta_{j_2'j}\delta_{j_3'1}
+(-1)^{h'} \delta_{j_2' 1} \delta_{j_3'j} \big) 
\delta_{h1}\delta_{j_1j}\delta_{j_2j}\delta_{j_3j}\delta_{j_1'1} \\+
\sqrt{\frac{2h+1}{3}} \sixj{j}{f}{1}{h}{j}{1}\big((-1)^j \delta_{j_2 j}\delta_{j_3 1}
+(-1)^{h} \delta_{j_2 1} \delta_{j_3 j} \big) 
\delta_{h'1}\delta_{j_1'j}\delta_{j_2'j}\delta_{j_3'j}\delta_{j_11}
\Bigg],
\end{array}
\ee
where 
\bea
B&=& \frac{j(j+1)(2j+1)}{4(g_1-g_2)j(j+1)+4g_2}, \cr
D&=& \sqrt{(2h+1)(2h'+1)}, \\
E(l)&=&\sum_x (2x+1) \sixj{f}{l}{x}{j}{1}{h}\sixj{f}{l}{x}{j}{1}{h'}\sixj{1}{j}{x}{1}{j}{j}.
\nonumber
\eea

\begin{figure}
\begin{center}
\includegraphics[width=70mm]{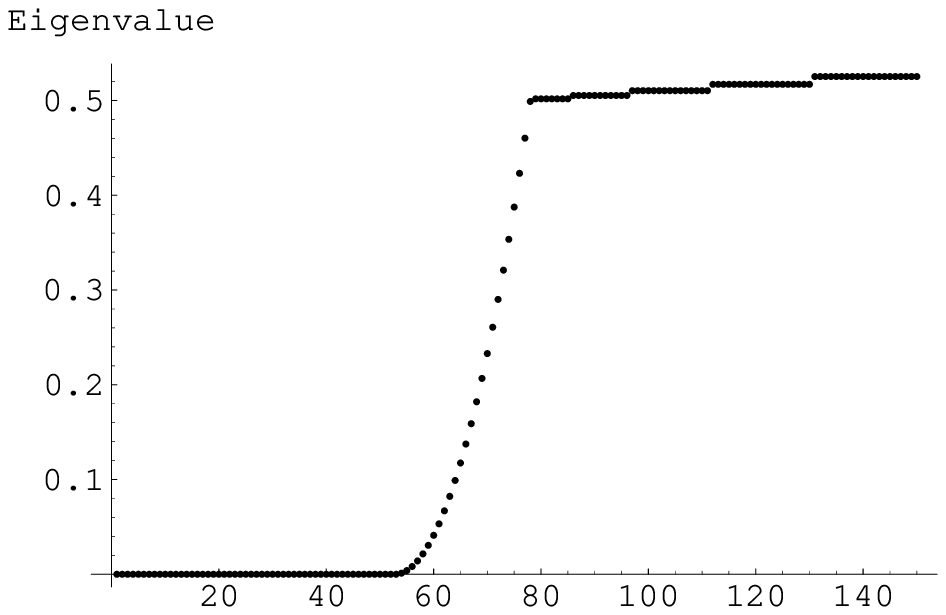}  
\includegraphics[width=70mm]{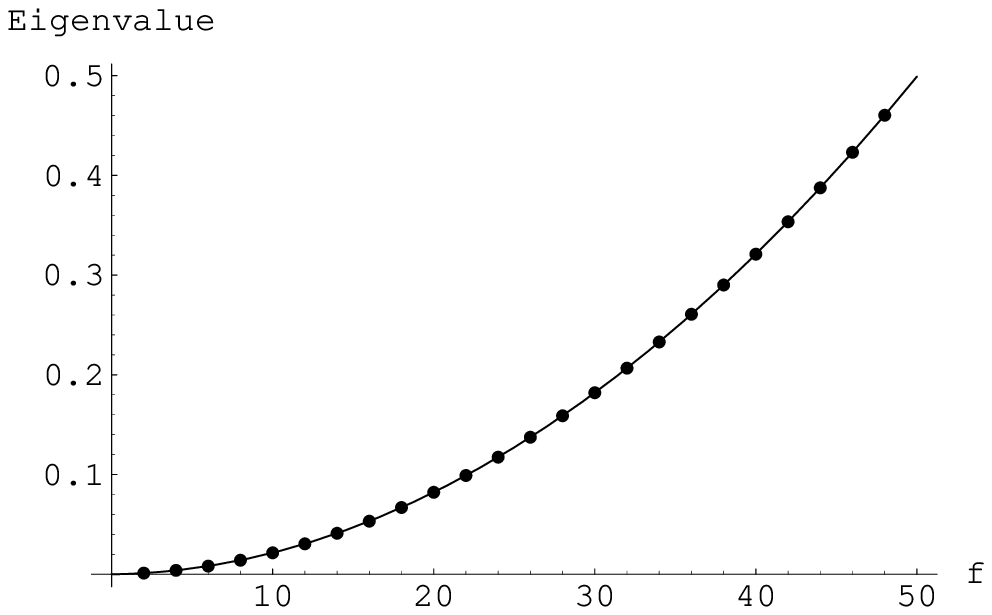}
\end{center}
\caption{The eigenvalues of $s_2(\alpha,\alpha',f)$ for $g_1=2,g_2=g_3=1,j=50$. 
In the left, the eigenvalues are
ordered in their values, of which the first 150 values are shown.  
There exist a series of eigenvalues which smoothly approach zero. 
In the right, the values of this series are well approximated by $0.000195723 f (f + 1)$.  }
\label{fig4}
\end{figure}

The analytical derivation of the eigenvalues of $s_2(\alpha,\alpha',f)$ seems to be hard.
I have instead analyzed the eigenvalues numerically. It is observed that the eigenvalues are non-negative
in a broad region of the parameters $g_{1,2,3},j$. Therefore the solutions are locally stable in that region. 
Another observation is concerning the zero modes.
There exists one zero eigenvalue for each $f=1,3,\cdots,2j-1$. Moreover, one additional zero eigenvalue exists
for each $f=j-1,j,j+1$. Since each eigenvalue has degeneracy $2f+1$, the total number of the zero modes is 
given by 
\be
\sum_{f=1,3,\cdots}^{2j-1} (2f+1)+ \sum_{f=j-1}^{j+1} (2f+1)=2j^2+7j+3.
\ee
This agrees with the number of the zero modes coming from the spontaneous breakdown of SO($n$) to
SO(3), where the total number of the indices is $n={\rm dim}(R)+{\rm dim}(G)=2j+1+3=2j+4$. 
Thus all the zero modes 
can be identified as the Nambu-Goldstone-like modes of SO($n$)$/$SO(3).
One also finds that there exist a series of eigenvalues which take rather smaller values  
and smoothly approach zero as shown in the left figure of Fig.\ref{fig4}.
These small values should characterize the effective continuum space generated by the solution. 
The distribution is such that one small value exists for each $f=2,4,\cdots \lesssim j$, and
can be well approximated by $const. f(f+1)$ as in the right figure of Fig.\ref{fig4}. These spectra agree
with those of the Laplacian on a two-sphere except the fact that $f=0$ and odd $f$ are lacking.
This implies that the SO(3) solution can be regarded as a fuzzy SO(3)/Z$_2$,
although explaining the lack of $f=0$ requires a detailed identification of the field associated with the 
spectra.

\section{Summary and comments}
\label{discussions}
In this talk, I have shown that the three-tensor models can be used as dynamical models of 
fuzzy spaces. The classical solutions to the models can be generally constructed from invariant tensors of 
Lie groups. 
Euclidean models which have actions bounded from below are introduced.
The explicit solutions with Lie group symmetries are found for a class of 
Euclidean model, and 
the SO(3) symmetric solution is identified as a fuzzy $S^2/Z_2$, 
based on the low-frequency spectra of the fluctuations around the solution.

A question is how many of all the classical solutions are accompanied with Lie group symmetries.  
If most of the solutions are, 
it would be interesting to use the three-tensor models as fuzzy higher dimensions 
\cite{Aschieri:2003vy} to explain the origin of the symmetries in our world.

As explained in Section \ref{introduction}, a motivation to consider the three-tensor models is 
to obtain a fuzzy analogue of general relativity. It will be especially interesting to 
investigate whether gravity-like low frequency modes exist in the models. 

%


\section*{Acknowledgments}
The author was supported by the Grant-in-Aid for Scientific Research No.13135213 and No.16540244
from the Ministry of Education, Science, Sports and Culture of Japan.


\end{document}